\documentclass[letter]{aa} 
\usepackage{graphicx}
\usepackage{txfonts}
\usepackage[round]{natbib}
\begin{document}
   \title{Probing MHD Shocks with high-$J$ CO observations: W28F}

   \author{A. Gusdorf
          \inst{1,2}
          \and
          S. Anderl\inst{3}
          \and
          R. G\"usten\inst{1}
           \and
          J. Stutzki\inst{4}
           \and
          H.-W. H\"ubers\inst{5,6}
          \and
          P. Hartogh\inst{7}
          \and
          S. Heyminck\inst{1}
          \and
          Y. Okada\inst{4}
          }

   \institute{Max Planck Institut f\"ur Radioastronomie, 
              Auf dem H\"ugel 69, 53121 Bonn, Germany, \email{agusdorf@mpifr-bonn.mpg.de}
            \and
            LERMA, UMR 8112 du CNRS, Observatoire de Paris, \'Ecole Normale Sup\'erieure, 24 rue Lhomond, F75231 Paris Cedex 05, France
            \and
             Argelander Institut f\"ur Astronomie, Universit\"at Bonn, Auf dem H\"ugel 71, 53121 Bonn, Germany
            \and
            I. Physikalisches Institut, Universit\"at zu K\"oln, Z\"ulpicher Strasse 77, 50937 K\"oln, Germany
            \and
            Deutsches Zentrum f\"ur Luft- und Raumfahrt, Institut f\"ur Planetenforschung, Rutherfordstrasse 2, 12489 Berlin, Germany
            \and
            Institut f\"ur Optik und Atomare Physik, Technische Universit\"at Berlin, Hardenbergstrasse 36, 10623 Berlin, Germany
            \and
            Max-Planck-Institut f\"ur Sonnensystemforschung, Max-Planck-Strasse 2, 37191 Katlenburg-Lindau, Germany
             }

   \date{Received September 15, 1996; accepted March 16, 1997}

 
  \abstract
   {Observing supernova remnants (SNRs) and modelling the shocks they are associated with is the best way to quantify the energy SNRs re-distribute back into the Interstellar Medium (ISM).}
   {We present comparisons of shock models with CO observations in the F knot of the W28 supernova remnant. These comparisons constitute a valuable tool to constrain both the shock characteristics and pre-shock conditions.}
   {New CO observations from the shocked regions with the APEX and SOFIA telescopes are presented and combined. The integrated intensities are compared to the outputs of a grid of models, which were combined from an MHD shock code that calculates the dynamical and chemical structure of these regions, and a radiative transfer module based on the \lq large velocity gradient' (LVG) approximation.}
   {We base our modelling method on the higher \textit{J} CO transitions, which unambiguously trace the passage of a shock wave. We provide fits for the blue- and red-lobe components of the observed shocks. We find that only stationary, C-type shock models can reproduce the observed levels of CO emission. Our best models are found for a pre-shock density of 10$^4$~cm$^{-3}$, with the magnetic field strength varying between 45 and 100~$\mu$G, and a higher shock velocity for the so-called blue shock ($\sim$25 km s$^{-1}$) than for the red one ($\sim$20 km s$^{-1}$). Our models also satisfactorily account for the pure rotational H$_2$ emission that is observed with \textit{Spitzer}.}
   {}

   \keywords{
   ISM: supernova remnants --
   ISM: individual objects: W28 --
   ISM: kinematics and dynamics --
   Physical data and processes: shock waves --
   Submillimeter: ISM --
   Infrared: ISM
               }

   \maketitle
%

\section{Introduction}

The interstellar medium (ISM) is in constant evolution, ruled by the energetic feedback from the cosmic cycle of star formation and stellar death. At the younger stages of star formation (bipolar outflows), and after the death of massive stars (SNRs), shock waves originating from the star interact with the ambient medium. They constitute an important mechanical energy input, and lead to the dispersion of molecular clouds and to the compression of cores, possibly triggering further star formation. Studying the signature of these interactions in the far-infrared and sub-mm range is paramount for understanding the physical and chemical conditions of the shocked regions and the large-scale roles of these feedback mechanisms.

Supernovae send shock waves through the ISM, where they successively carve out large hot and ionised cavities. They subsequently emit strong line radiations (optical/UV), and eventually interact with molecular clouds, driving lower-velocity shocks. Similar to their bipolar outflow equivalents, these shocks heat, compress, and accelerate the ambient medium before cooling down through molecular emission (\citealt{Vandishoeck93}, \citealt{Yuan11}, hereafter Y11).

Valuable information has been provided by ISO~\citep{Cesarsky99, Snell05} and \textit{Spitzer} (\citealt{Neufeld07}, hereafter N07), but neither of those instruments provided sufficient spectral resolution to allow for a detailed study of the shock mechanisms. High-$J$ CO emission is one of the most interesting diagnostics of SNRs. CO is indeed a stable and abundant molecule, and an important contributor to the cooling of these regions, whose high-frequency emission is expected to be a \lq pure' shock tracer. Observations of the latter must be carried out from above the Earth's atmosphere. As part of a multi-wavelength study of MHD shocks that also includes \textit{Herschel} data, we present here the first velocity-resolved CO (11--10) observations towards a prominent SNR-driven shock with the GREAT spectrometer onboard SOFIA, and combine them with new lower-$J$ ones in a shock-model analysis.

\section{The supernova remnant W28}

W28 is an old ($>$10$^{4.5}$ yr, \citealt{Claussen99}) SNR in its radiative phase of evolution, with a non-thermal radio shell centrally filled with thermal X-ray emission. Lying in a complex region of the Galactic disk at a distance of 1.9$\pm$0.3~kpc~\citep{Velazquez02}, its structure in the 327 MHz radio continuum represents a bubble-like shape of about 40$\times$30~pc~\citep{Frail93}. Early on, molecular line emission peaks, not associated with star formation activity, but revealing broad lines, were suggested as evidence for interaction of the remnant with surrounding molecular clouds~\citep{Wootten81}. Later studies spatially resolved the shocked CO gas layers from the ambient gas \citep{Frail98,Arikawa99}. OH maser spots line up with the post-shock gas layers \citep{Frail94,Claussen97,Hoffman05}, for which the strongest masers VLBA polarisation studies yield line-of-sight magnetic field strengths of up to 2 mG \citep{Claussen99}. Pure rotational transitions of H$_2$ have been detected with ISO~\citep{Reach00} and were more recently observed with \textit{Spitzer}, better resolved spatially and spectrally,  by N07 and Y11. 

Recently, very high energy (TeV) $\gamma$-ray emission has been detected by HESS \citep{Aharonian08}, Fermi \citep{Abdo10}, and AGILE \citep{Giuliani10}, spatially slightly extended and coincident with the bright interaction zones, W28-E and -F. If interpreted as the result of hadronic cosmic ray interactions in the dense gas ($\pi^0$ decay), a cosmic ray density enhancement by an order of magnitude is required (which is supplied/accelerated by the SNR).

\subsection{Sub-mm CO observations of W28F}
\label{sub:opcooow}

APEX\footnote{This publication is partly based on data acquired with the Atacama Pathfinder EXperiment (APEX). APEX is a collaboration between the Max-Planck-Institut f\"ur Radioastronomie, the European Southern Observatory, and the Onsala Space Observatory.} \citep{Guesten06} observations towards W28F were conducted in 2009 and will be the subject of a forthcoming publication (Gusdorf et al., in prep.). For the present study, we used 100$'' \times$100$''$ maps in the $^{13}$CO (3--2), CO (3--2), (4--3), (6--5), and (7--6) transitions, described in Appendix~\ref{sec:tao}.

   \begin{figure}
   \centering
   \includegraphics[width=9cm]{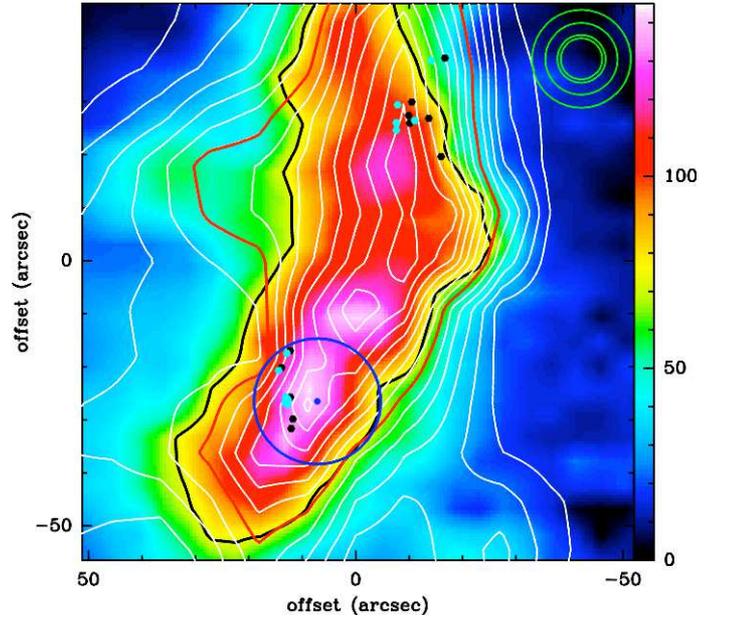}
      \caption{Overlay of the velocity-integrated CO (6--5) (colour background) with the CO (3--2) (white contours) emission observed with the APEX telescope. For both lines, the intensity was integrated between -30 and 40 km s$^{-1}$. The wedge unit is K km s$^{-1}$ in antenna temperature. The CO (3--2) contours are from 30 to 160~$\sigma$, in steps of 10$\sigma$ = 16 K km s$^{-1}$. The half-maximum contours of the CO (3--2) and (6--5) maps are indicated in red and black, respectively. The dark blue circle indicates the position and beam size of the SOFIA/GREAT observations. The APEX beam sizes of our CO (3--2), (4--3), (6--5), and (7--6) observations are also provided (upper right corner light green circles, see also Table~\ref{tablea1}). The maps are centred at (R.A.$_{[\rm{J}2000]}$=$18^h01^m52\fs3$, Dec$_{[\rm{J}2000]}$=$-23^\circ19'$25$''$). The black and light blue hexagons mark the position of the OH masers observed by \citet{Claussen97} and \citet{Hoffman05}.}
         \label{figure1}
   \end{figure}

In Fig.~\ref{figure1} the velocity-integrated CO (6--5) broad-line emission of W28F is shown overlaid with the CO (3--2) emission (white contours): a north-south elongated structure of about 100$''$ height and 30$''$ width traces the same warm accelerated post-shocked gas. In our high-resolution CO (6--5) data the structure is resolved, though probably still sub-structured similar to what is seen in H$_2$, e.g., Y11. Comparison with the distributions of excited H$_2$ and OH masers (whose locations also mark the leading edge of the non-thermal radio shell) suggests a textbook morphology of an SNR-molecular cloud interaction: the shock propagates E-NE into the ambient cloud that extends east for several arcmins. Hot H$_2$ and OH masers mark the first signposts of the shock-compressed gas. Farther downstream, the gas cooling is seen prominently in warm CO. The shock impact appears edge-on, but the fact that high - projected - streaming velocities are indeed observed (-30 km s$^{-1}$ with respect to the ambient cloud) requires a significant inclination angle. 
   
   \begin{figure}
   \centering
   \includegraphics[width=9cm]{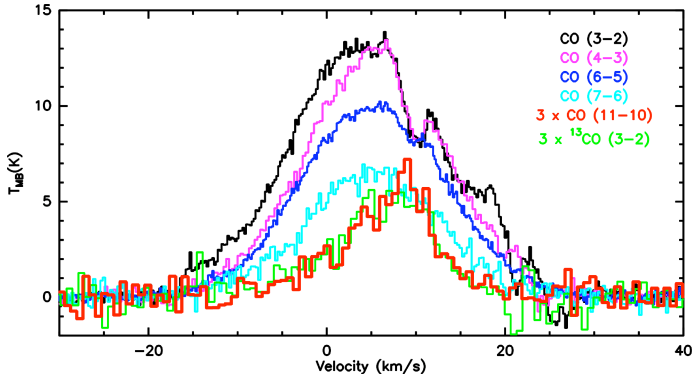} 
      \caption{CO transitions observed in the position (+7$''$,-26$''$) indicated in Fig.~\ref{figure1}: APEX (3--2), black (corresponding $^{13}$CO, green); (4--3), pink; (6--5), dark blue; (7--6), light blue; and SOFIA (11--10), red. The $^{13}$CO (3--2) and CO (11--10) profiles were multiplied by three for comparison purposes. Respective CO spectral resolutions are 0.212, 0.318, 0.159, 0.272, and 0.693 km s$^{-1}$, and 0.664 km s$^{-1}$ for $^{13}$CO (3--2).     
              }
         \label{figure2}
   \end{figure}   
   
We have selected the most prominent position in the southern extension of W28F, marked with the blue circle in Fig.1, for our shock analysis. This position was also covered by \textit{Spitzer}, offering a set of complementary H$_2$ data. Fig.~\ref{figure2} shows the APEX spectra obtained towards this position, in main beam temperature units, all convolved to the 23.7$''$ beam of the SOFIA observations (Sect.~\ref{sub:oadr}). Absorption notwithstanding, the spectra show Gaussian profiles, with all line wings extending in blue (-30~km~s$^{-1}$) and red (+15~km~s$^{-1}$) velocities with respect to the ambient cloud ($\sim$10~km ~s$^{-1}$). The higher the line frequency, the narrower the line profile, with a typical linewidth of about 20 km s$^{-1}$ for CO (7--6). We detect red-shifted line-of-sight absorption features in all lines up to CO (6--5), the deepest arising at the ambient cloud velocity. Off-position contamination results in minor absorption features at 20 and 25 km s$^{-1}$ in the (3--2) and (4--3) profiles. Comparison with the $^{13}$CO(3-2) profile is also shown in Fig.~\ref{figure2}. An analysis of the line temperature ratio of  $^{12}$CO/$^{13}$CO yields an optical thickness value of 3--7 in the wings of the $^{12}$CO (3--2), assuming a typical interstellar abundance ratio of 50--60 (e.g., \citealt{Langer93}).

\subsection{Far-infrared CO spectroscopy with GREAT/SOFIA}
\label{sub:oadr}

The observations towards W28F were conducted with the GREAT\footnote{GREAT is a development by the MPI f\"ur Radioastronomie and the KOSMA$/$Universit\"at zu K\"oln, in cooperation with the MPI f\"ur Sonnensystemforschung and the DLR Institut f\"ur Planetenforschung.} spectrometer (\citealt{Heyminck12}) during SOFIA's flight from Stuttgart to Washington on September 21 2011. Only one position could be observed, towards the southern tip of the shocked cloud at offset (+7$''$,-26$''$) (Fig.~\ref{figure1}). The CO (11--10) line was tuned to the frequency 1267.015 GHz LSB. The receiver was connected to a digital FFT spectrometer (\citealt{Klein12}) providing a bandwidth of 1.5 GHz with a spectral resolution of 0.05 km s$^{-1}$. The observations were performed in double beam-switching mode, with an amplitude of 80$''$ (or a throw of 160$''$) at the position angle of 135$^\circ$ (NE--SW), and a phase time of 0.5 sec. The nominal focus position was updated regularly against temperature drifts of the telescope structure. The pointing was established with the optical guide cameras to an accuracy of $\sim$5$''$. The beam width $\Theta_{\rm{mb}}$ is 23.7$''$; the main beam and forward efficiencies are $\eta_{\rm{mb}}$~=~0.54 and $\eta_{\rm f}$~=~0.95. The integration time was 13 min ON source, for a final r.m.s of 0.66~K. The data were calibrated with the KOSMA/GREAT calibrator (\citealt{Guan12}), removing residual telluric lines, and subsequently processed with the CLASS software\footnote{http://www.iram.fr/IRAMFR/GILDAS}.

The CO (11--10) spectrum, overlaid on the sub-mm lines in Fig.~\ref{figure2}, reveals a markedly different profile: weak emission only is seen from the high-velocity gas, while the line is more prominent at low velocities. The profile basically follows the shape of the optically thin $^{13}$CO(3-2).

\section{Discussion}
\label{sec:dis}

\subsection{The observations}
\label{sub:theobs}

Although most likely part of a single original shock clump, we separated the profiles into a blue lobe (-30 to 7.5-12.5 km s$^{-1}$) and a red lobe (7.5-12.5 to 40 km s$^{-1}$), and fited the data independently. The uncertainty of the upper (lower) limit of those ranges reflects our lack of knowledge of the ambient velocity component. In Fig.~\ref{figure3} we plot the velocity-integrated intensities of the CO lines, all convolved to the same angular resolution of 23.7$''$, against the rotational quantum number of their upper level in a so-called \lq spectral line energy distribution' (or SLED, filled black squares with errorbars). The underlying assumption is that the filling factor is the same for all CO lines, which is validated by the similarity between their emitting regions (see for instance Fig.~\ref{figure1}, where the half-maximum contours of CO (3--2) and (6--5) coincide at the available resolution). For Fig.~\ref{figure3}, the assumption was made that all observed transitions present a circular emission region of radius 25$''$, corresponding to a filling factor of 0.53 (compatible with our maps, see Fig.~\ref{figure1}).

The upper (lower) panel shows the diagram associated with the blue (red) shock component. The lower limit to the integrated intensity corresponds to the integration of the observed profile in the smallest velocity range, whereas upper limits were obtained by integrating Gaussian fits to the observation on the largest velocity range. The fits were adjusted to recover the shock flux lost through absorption and based on the un-absorbed parts of the profiles. Although yielding high errorbars on our measurements, specially for low $J_{\rm{up}}$ values, this method also provides the most conservative approach to our blue-red decomposition of the CO emission.

\subsection{The models}
\label{sub:themod}

We then compared the resulting integrated intensity diagrams to modelled ones. To build those, we used a radiative transfer module based on the \lq large velocity gradient' (LVG) approximation to characterise the emission from the CO molecule over outputs generated by a state-of-the-art model that calculates the structure of one-dimension, stationary shock layers (or approximations of non-stationary layers). This method has already been used and extensively introduced in~\citet{Gusdorf082}. Since then, the LVG module has been modified to incorporate the latest collisional rate coefficients of CO with H$_2$ computed by~\citet{Yang10}, but the shock model is the same. Based on a set of input parameters (pre-shock density, shock velocity, type, and age, and magnetic field parameter value $b$ such as B[$\mu$G]~=~$b~\times~\sqrt{n_{\rm H}~[{\rm cm}^{-3}]}$), it calculates the structure of a shock layer, providing dynamical (velocity, density, temperature), and chemical (fractional abundances of more than 125 species linked by over 1000 reactions) variables values at each point of the layer. The relevant outputs are then used by our LVG module, which calculates the level population and line emissivities of the considered molecule. In the present case, the line temperatures were computed for transitions of CO up to (40--39) at each point of the shock layer, over which they were then integrated to form our model's SLED. Because of the importance of H$_2$ (key gas coolant, abundant molecule), its radiative transfer was treated within the shock model. Its abundance was calculated at each point of the shock layer, based on the processes listed in~\citet{Lebourlot02}, and the populations of the 150 first levels were also calculated inside the shock code, their contribution to the cooling being included in the dynamical calculations. It is hence also possible to associate an H$_2$ excitation diagram to each model of the grid, see Sect.~\ref{sub:excdia}.

Our model grid consists of a large sample of integrated intensity diagrams, obtained for stationary C- and J-type, and non-stationary CJ-type shocks. It covers four pre-shock densities (from 10$^3$ to 10$^6$ cm$^{-3}$), $0.45 \leq b \leq 2$, over a range of velocities spanning from 5 km s$^{-1}$ up to the maximum, \lq critical' value above which a C-shock can no longer be maintained, which depends on the other parameters. We then independently compared the observed blue- and the red-shock component to the whole grid diagrams. We used a $\chi^2$ routine that was set up to provide the best fits to our purest shock-tracing lines: CO (6--5), (7--6), and (11--10), to reduce any effect induced by ambient emission contamination on the lower-lying transitions. 

\subsection{The results}
\label{sub:theres}

   \begin{figure}
   \centering
   \includegraphics[width=0.3\textwidth]{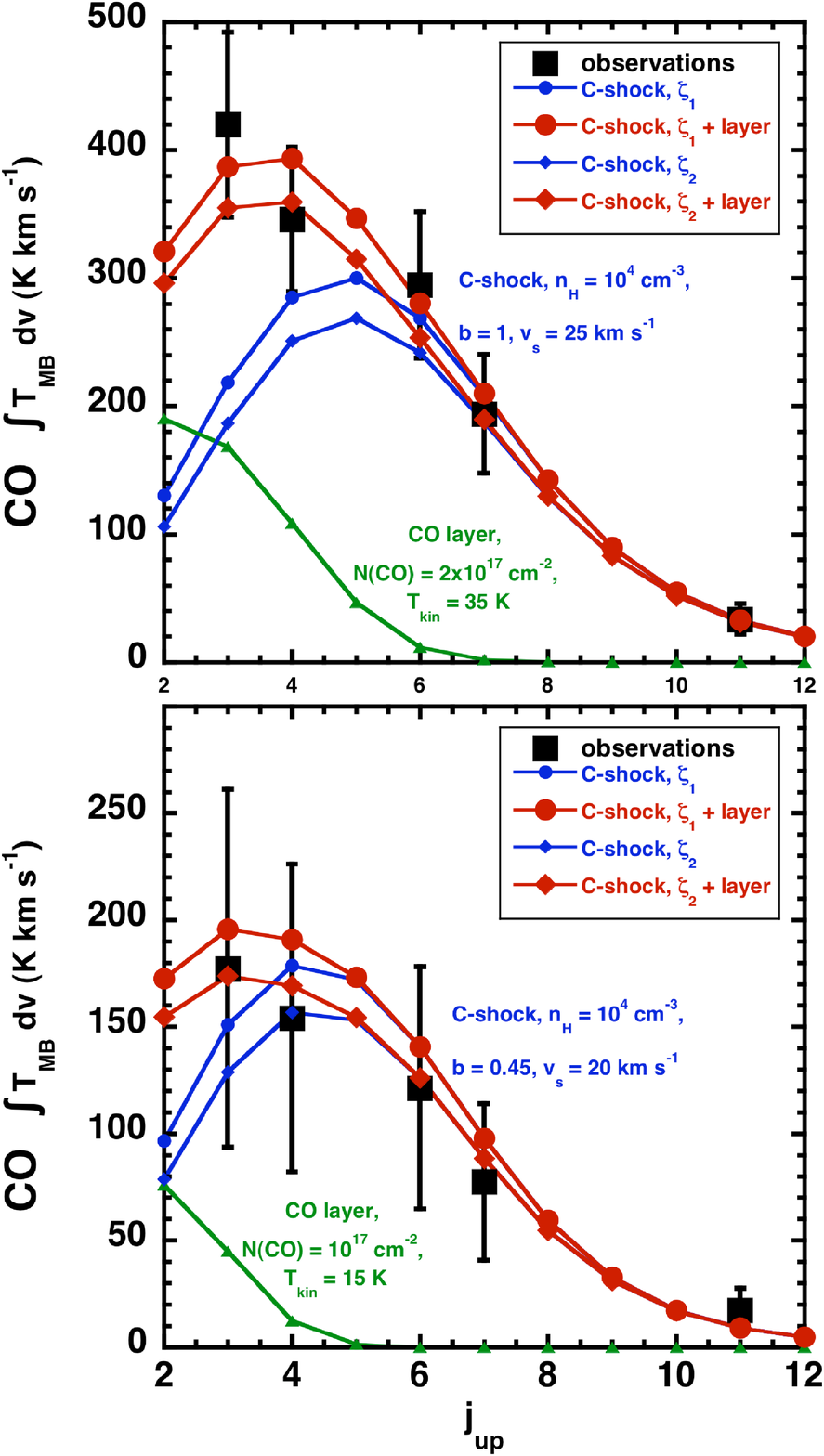}
      \caption{Best-model comparisons between CO observations and models for each observed shock: the blue- (upper panel) and the red- (lower panel) components. Observations are marked by the black squares, our individual best-shock models are in blue line and circles ($\zeta_1$) or diamonds ($\zeta_2$), the warm layer that we used to compensate the ambient emission affecting the (3--2) and (4--3) transitions in green line and triangles, and the sum of each of these components is represented by a red line and circles or diamonds.}
         \label{figure3}
   \end{figure}

The upper (lower) panel of Fig.~\ref{figure3} shows our best-fitting models for the blue (red) component. We found that only stationary, C-type shock models can reproduce the levels of observed CO (6--5), (7--6) and (11-10) integrated intensities. In both cases, the pre-shock density is 10$^4$~cm$^{-3}$, for which a stationary state is typically reached for a shock age of 10$^4$ years, in agreement with the age of the remnant of 3.5$\times 10^4$ years quoted by \citet{Giuliani10}. The modelled shock velocities, 25 and 20 km s$^{-1}$, respectively, are rough upper limits to the observed ones, an expected conclusion given the one-dimensional nature of our models. Additionally, our models provide respective constraints to the magnetic field component perpendicular to the shock layers, with parameter $b$ of 1 and 0.45. The difference between those two values can be explained by a projection effect. In both cases, the direct post-shock density is of the order of 2-3$\times 10^5$cm$^{-3}$, yielding expected post-shock values for this component of 0.05-0.55 mG, compatible with those inferred by \citet{Claussen99} (2 mG for the total post-shock magnetic field), and by \citet{Hoffman05} (0.3--1.1 mG based on the OH maser measurements in the positions indicated in Fig.~\ref{figure1} located within the blue beam). Finally, our models produce typical OH column densities of a few 10$^{16}$cm$^{-2}$. This complies with the list of requirements set out by \citet{Lockett99} for exciting the observed OH 1720 MHz masers, together with the C- nature of the considered shocks, as well as the temperature and density conditions they create.

The respective CO column density found in each of the red and blue shock models is 11 and 5$\times 10^{17}$cm$^{-3}$. However, our shock models failed to account for the important observed levels of CO (3--2) and (4--3) emission. We found it possible to model this lower-$J$ surplus emission by adding a thin layer of gas to our shock model results. Its corresponding emission was calculated with our LVG module used in a \lq homogeneous slab' mode, and indeed generated significant emission only from the (3--2) and (4--3) lines, as can also be seen in Fig.~\ref{figure3}. We used a linewidth of 10 km~s$^{-1}$ and a density of $n_{\rm H} = 10^4$~cm$^{-3}$ with a fractional abundance of $X(\rm{CO}) = 10^{-4}$ for the blue and red cases. We found that this \lq CO layer' is warm (35~K and 15~K for the blue and red components, respectively), and adds to the pure-shock CO column density in both components, from 2$\times$10$^{17}$ cm$^{-2}$ for the blue component, and 10$^{17}$ cm$^{-2}$ for the red one. Our final result, the sum of the C-shock and \lq CO layer', is also shown in Fig.~\ref{figure3}. 

To understand the physical origin of this emitting layer, we varied the cosmic ray ionisation rate in our best-fitting models from its solar value $\zeta_1 = 5 \times 10^{-17}$ s$^{-1}$ to the $\zeta_2 = 3.4 \times 10^{-16}$ s$^{-1}$ indicated by~\citet{Hewitt09} in the specific case of W28. In our modelling, this modification impacts on the chemistry (higher ionisation fraction) and the physics (warmer gas) of our outputs. The resulting shock profile is more narrow (hence generating slightly less CO emission, see Fig.~\ref{figure3}), with higher post-shock temperatures close to the aforementioned CO layer. Because one of the limitations of our shock model is the absence of significant velocity gradient in the post-shock region, the warm CO layer component might correspond to a post-shock layer of gas mildly heated by the energetic radiation that seems to exist in the region (UV or more energetic, as characterised by~\citealt{Hewitt09}). The rigorous inclusion of these radiative effects in our shock model is still in progress. In Sect.~\ref{sec:tso} we present a consistency check based on the use of the \textit{Spitzer} H$_2$ observations.

\begin{acknowledgements}
      We thank the SOFIA engineering and operations teams, whose support has been essential during basic science flights, and the DSI telescope engineering team. Based [in part] on observations made with the NASA/DLR Stratospheric Observatory for Infrared Astronomy. SOFIA Science Mission Operations are conducted jointly by the Universities Space Research Association, Inc., under NASA contract NAS2-97001, and the Deutsches SOFIA Institut under DLR contract 50 OK 0901. S. Anderl acknowledges support by the Deutsche Forschungsgemeinschaft within the SFB 956 "Conditions and impact of star formation", the International Max Planck Research School (IMPRS) for Radio and Infrared Astronomy at the Universities of Bonn and Cologne, and the Bonn-Cologne Graduate School of Physics and Astronomy. A. Gusdorf acknowledges support by the grant ANR-09-BLAN-0231-01 from the French
 {\it Agence Nationale de la Recherche} as part of the SCHISM project.     
\end{acknowledgements}



\bibliographystyle{aa}
\bibliography{biblio}

\Online

\begin{appendix} 
\section{The APEX observations}
\label{sec:tao}

\begin{figure}
   \centering
   \includegraphics[width=9cm]{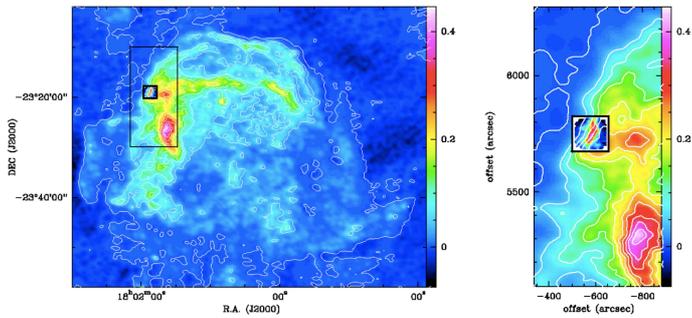}
      \caption{Location of the field covered by our CO observations on the larger-scale radio continuum image at 327 MHz taken from \citet{Claussen97}: entire SNR in the left panel, zoom in the right panel. The shown CO observations are the (6--5) map, also displayed in Fig.~\ref{figure1}.
              }
         \label{figuresup}
   \end{figure}

APEX observations towards the supernova remnant W28F were conducted in several runs in the year 2009 (in May, June, August and October). We used of a great part of the suite of heterodyne receivers available for this facility: APEX--2 \citet{Risacher06}, FLASH460 \citet{Heyminck06}, and CHAMP$^+$ \citet{Kasemann06,Guesten08}, in combination with the MPIfR fast Fourier transform spectrometer backend (FFTS, \citealt{Klein06}). The central position of all observations was set to be $\alpha_{[\rm{J}2000]}$=$18^h01^m51\fs78$, $\beta_{[\rm{J}2000]}$=$-23^\circ18'58$\farcs$50$. Focus was checked at the beginning of each observing session, after sunrise and/or sunset on Mars, or on Jupiter. Line and continuum pointing was locally checked on RAFGL1922, G10.47B1, NGC6334-I or SgrB2(N). The pointing accuracy was found to be of the order of 5$''$ rms, regardless of the receiver that was used. Table~\ref{tablea1} contains the main characteristics of the observed lines and corresponding observing set-ups: frequency, beam size, sampling, used receiver, observing days, forward and beam efficiency, system temperature, spectral resolution, and finally the velocity interval that was used to generate the integrated intensity maps. The observations were performed in position-switching/raster mode using the APECS software \citet{Muders06}. The data were reduced with the CLASS software (see http://www.iram.fr/IRAMFR/GILDAS). For all observations, the maximum number of channels available in the backend was used (8192), except for CO (4--3), for which only 2048 channels were used, leading to the spectral resolutions indicated in Table~\ref{tablea1}. Maps were obtained for all considered transitions, covering the field introduced in Fig.~\ref{figure1} and put in the perspective of the whole SNR in Fig.~\ref{figuresup}.

\begin{table*}
\caption{Observed lines and corresponding telescope parameters for the APEX observations of W28F.}             
\label{tablea1}      
\centering                          
\begin{tabular}{c  c  c  c  c c }        
\hline           
line & CO (3--2) & CO (4--3) & CO (6--5) & CO (7--6) & $^{13}$CO (3--2)  \\
\hline
\hline
\footnotesize{$\nu$ (GHz)} & \footnotesize{345.796} & \footnotesize{461.041} & \footnotesize{691.473} & \footnotesize{806.652} & \footnotesize{330.588} \\
\footnotesize{FWHM ($''$)} & \footnotesize{18.1} & \footnotesize{13.5} & \footnotesize{9.0} & \footnotesize{7.7} & \footnotesize{18.9} \\
\footnotesize{sampling ($''$)} & \footnotesize{10} & \footnotesize{7} & \footnotesize{4} & \footnotesize{4} & \footnotesize{10} \\
\hline
\footnotesize{receiver} & \footnotesize{HET345} & \footnotesize{FLASH460} & \footnotesize{CHAMP$^+$} & \footnotesize{CHAMP$^+$} & \footnotesize{HET345} \\
\footnotesize{observing days} & \footnotesize{04-05/08} & \footnotesize{04/06} & \footnotesize{13-14-15/08} & \footnotesize{13-14-15/08} & \footnotesize{05-06-07/08} \\
\hline
\footnotesize{$F_{ \rm eff}$} & \footnotesize{0.97} & \footnotesize{0.95} & \footnotesize{0.95} & \footnotesize{0.95} & \footnotesize{0.97} \\
\footnotesize{$B_{ \rm eff}$} & \footnotesize{0.73} & \footnotesize{0.60} & \footnotesize{0.52} & \footnotesize{0.49} & \footnotesize{0.73} \\
\hline 
\footnotesize{$T_{\rm sys}$ (K)} & \footnotesize{279--288} & \footnotesize{372--439} & \footnotesize{1330--1722} & \footnotesize{3524--5850} & \footnotesize{357--377} \\
\footnotesize{$\Delta \varv$ (km s$^{-1}$)} & \footnotesize{0.106} & \footnotesize{0.318} & \footnotesize{0.079} & \footnotesize{0.068} & \footnotesize{0.166} \\    
\hline                                  
\footnotesize{reference offset ($''$)} & \footnotesize{(-100,-50)} & \footnotesize{(-120,0)}\tablefootmark{a} & \footnotesize{(-100,-50)} & \footnotesize{(-100,-50)} & \footnotesize{(-100,-50)}\tablefootmark{b} \\
\hline
\footnotesize{$\varv$ interval (km s$^{-1}$)} & \footnotesize{-30/40} & \footnotesize{-30/40} & \footnotesize{-30/40} & \footnotesize{-10/25} & \footnotesize{-30/40} \\ 
\hline
\end{tabular}
\tablefoot{
\tablefoottext{a}{The observations were carried out with this off position, and were corrected a posteriori from its potential contamination, by means of an off spectrum on the (-120,0) position taken with an off position at (-240,0).}\\
\tablefoottext{b}{The observations were carried out with this off position, and were corrected a posteriori from its potential contamination, by means of an off spectrum on the (-100,-50) position taken with an off position at (1000,1000).} 
}
\end{table*}

\section{The H$_2$ observations}
\label{sec:tso}

\subsection{The dataset}

   \begin{figure}
   \centering
   \includegraphics[width=6cm]{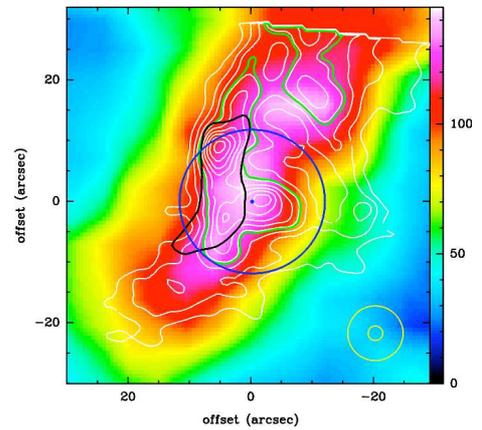}
      \caption{Overlay of the map of CO (6--5) emission observed by the APEX telescope (colour background) with the H$_2$ 0--0 S(5) emission (white contours), observed with the \textit{Spitzer} telescope. The wedge unit is K km s$^{-1}$ (antenna temperature) and refers to the CO observations. The H$_2$~0--0~S(5) contours are from 50 to 210~$\sigma$, in steps of $20 \sigma \simeq 1.6 \times 10^{-4}$ erg cm$^{-2}$ s$^{-1}$ sr$^{-1}$. The green contour defines the half-maximum contour of this transition. Like in Fig.~\ref{figure1}, the blue circle indicates the beam size of the SOFIA/GREAT observations, on the position of the centre of the circle, marked by a blue dot. The beam and pixel sizes of the CO (6--5), and H$_2$~0--0~S(5) observations are also provided in yellow (big and small, respectively) circles in the lower right corner. The black contour delineates the half-maximum contour of the H$_2$~0--0~S(2) transition. The field is smaller than in Fig.~\ref{figure1}, and the (0,0) position is that of the \textit{Spitzer} observations (see Appendix~\ref{sec:tso}).      
              }
         \label{figurea1}
   \end{figure}
   
As a consistency check for our models, we used of the \textit{Spitzer}/IRS observations of the H$_2$ pure rotational transitions (0--0~S(0) up to S(7)), reported and analysed in N07 and Y11. Although the original dataset also includes other ionised species, we chose to use only the H$_2$ data. The raw product communicated to us by David Neufeld contains rotational transitions maps, with 1.2$''$ per pixel, centred on $\alpha_{[\rm{J}2000]}$=$18^h01^m52\fs32$, $\beta_{[\rm{J}2000]}$=$-23^\circ19'24$\farcs$92$. Fig.~\ref{figurea1} shows an overlay of our APEX CO~(6--5) map with the H$_2$~0--0~S(5) region observed by \textit{Spitzer}. The figure shows coinciding maxima between the two datasets in the selected position, and a slightly different emission distribution. This might be the effect of the better spatial resolution of the H$_2$ data, which reveal more peaks than in CO. This overlay also shows the slightly different morphology of the emission of the S(2) (half maximum contour in black) and S(5) (half maximum contour in green) transitions, at the available resolution.

\subsection{Excitation diagram}
\label{sub:excdia}

We performed a consistency check on our modelling (see Sect.~\ref{sec:dis}) using the excitation diagram derived for the selected emission region. The H$_2$ excitation diagram displays ln($N_{\varv j}/g_j$) as a function of $E_{\varv j}/k_{\rm B}$, where $N_{\varv j}$ (cm$^{-2}$) is the column density of the rovibrational level ($v, J$), $E_{\varv j}/k_{\rm B}$ is its excitation energy (in K), and $g_j = (2j+1)(2I+1)$ its statistical weight (with $I=1$ and $I=0$ in the respective cases of ortho- and para-H$_2$). If the gas is thermalised at a single temperature, all points in the diagram fall on a straight~line. The selected position for the present study was introduced in Sect.~\ref{sub:opcooow}, and can be seen in Figs.~\ref{figure1} and \ref{figurea1}. The column density of the higher level of each considered transition was extracted by averaging the line intensities in a 11.85$''$ radius circular region, consistent with our handling of the CO data. In the process, we corrected the line intensities for extinction, adopting the visual extinction values from N07, $A_{\rm \varv} = 3-4$, and using the interstellar extinction law of~\citet{Rieke85}. The resulting excitation diagram can be seen in Fig.~\ref{figurea2}. We note that the intensity of the H$_2$ 0--0 S(6) transition ($J = 8$) cannot be determined reliably, because the line is blended with a strong 6.2~$\mu$m PAH feature.       

\subsection{Comparisons with our models}

\begin{figure}
   \centering
   \includegraphics[width=9cm]{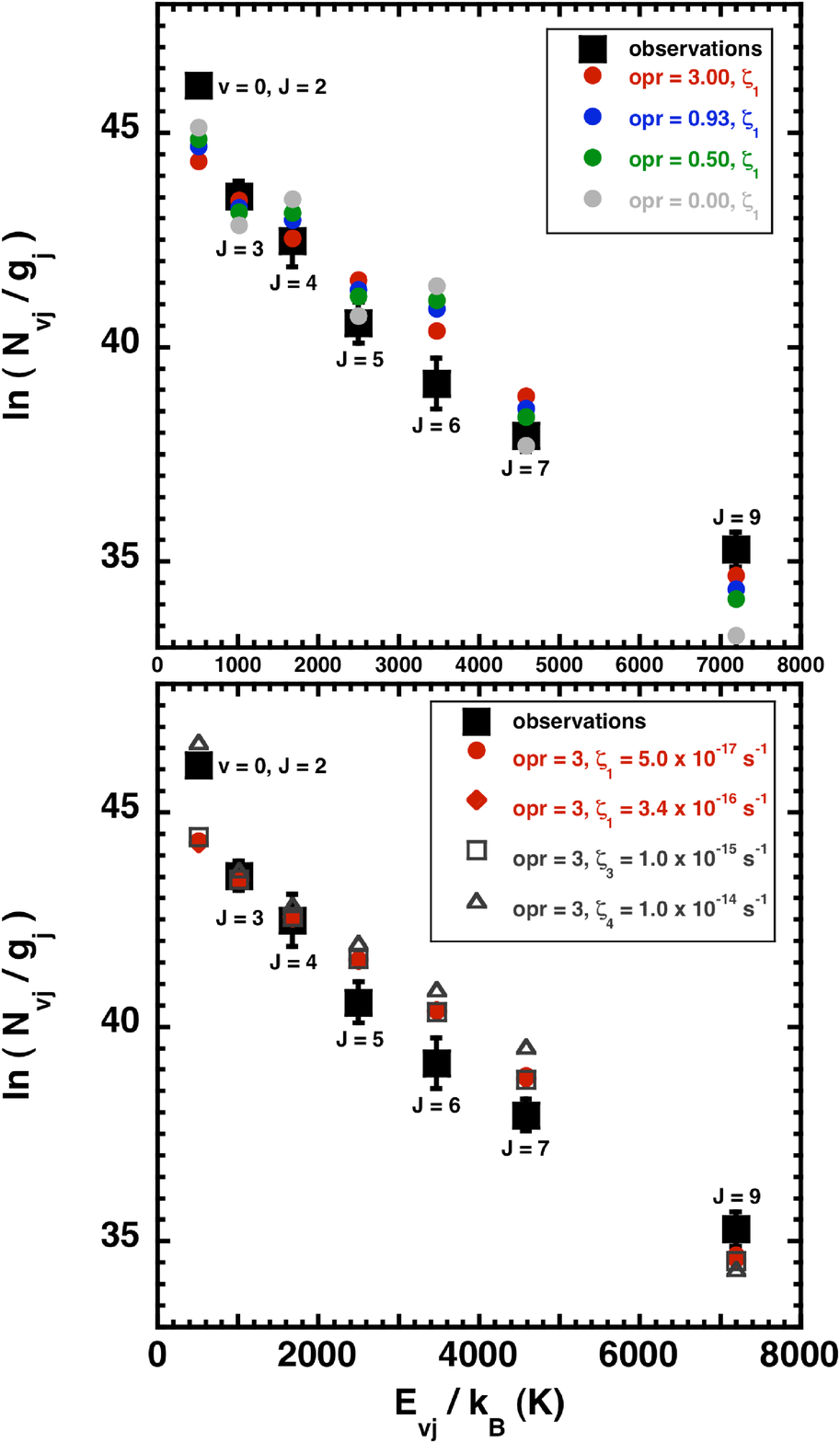}
      \caption{H$_2$ excitation diagram comparisons. \textit{Upper panel:} evolution of the modelled excitation diagram obtained for $\zeta_1 = 5 \times 10^{-17}$ s$^{-1}$, varying the initial value of the OPR from the unrealistic, extreme-case 0 value to its equilibrium one, 3. \textit{Lower panel:} influence of the cosmic ray ionisation rate variation on our modelled excitation diagram, with the initial OPR set to 3.}
         \label{figurea2}
   \end{figure}

Unlike our CO observations, the \textit{Spitzer}/IRS observations are not spectrally resolved (owing to the relatively low resolving power of the short low module in the range 60--130). On the other hand, given the minimum energy of the levels excited in the H$_2$ transitions (for 0--0~S(0), $E_{\rm u} \sim 509.9$~K), we can make the double assumption that the lines are not contaminated by ambient emission or self-absorption, and that the measured line intensities are the result from the blue- and red-shock present in the line of sight. Therefore, one must compare the observed H$_2$ level populations to what is generated by the sum of our best-fitting blue- and red- shock models. Fig.~\ref{figurea2} shows the comparison between the excitation diagram derived from the sum of our two CO best-fitting models and the observed one. The H$_2$ excitation diagram comprising the sum of our CO best-fitting models is shown in both panels in red circles, and provides a satisfying fit to the observations. With the aim of improving the quality of this fit, we also studied the influence of the initial ortho-to-para ratio (OPR) value, upper panel, and of the cosmic ray ionisation rate value, lower panel.

In the upper panel, for a cosmic ray ionisation rate value adopted as the solar one, $\zeta_1 = 5 \times 10^{-17}$ s$^{-1}$, the influence of the initial OPR value is studied. In their models, N07 inferred a value of 0.93 for the \lq warm' component ($\sim$322 K). On the other hand, the value associated to their \lq hot' ($\sim$1040 K) component could not be estimated, owing to the large uncertainty associated to the determination of the H$_2$ 0--0 S(6) flux. In our shock models, the OPR value is consistently calculated at each point of the shocked layer, and is mostly the result of conversion reactions between H$_2$ and H, H$^+$ or H$_3^+$. The heating associated to the passage of the wave increases the OPR value towards the equilibrium value of 3.0. Nevertheless, this value is reached only when it is also the initial one. The N07 situation is then probably adequately approximated in our models where the OPR initial value is less than 3. However, the influence of this parameter is minimal on the excitation diagram, where it only seems to create a slight saw-tooth pattern between the odd and even values of $J$.


We finally investigated the effects of the high-energy radiation field on the excitation diagram in the lower panel of Fig.~\ref{figurea2}. The limits of our modelling are reached because the only handle that we have to study this effect on our outputs is the variation of the cosmic ray ionisation rate value, which affects the corresponding chemistry (higher ionisation fraction) and physics (warmer gas). A proper treatment of the energetic radiation components listed in \citet{Hewitt09} should indeed incorporate UV pumping or self-shielding, as would be the case in PDR regions (e.g., \citealt{Habart11}), as well as chemical and H$_2$ excitation effects by X-ray (see \citealt{Dalgarno99}) and  cosmic rays (e.g., \citealt{Ferland08}). Keeping these limits in mind, we found that even strong modifications of the cosmic ray ionisation rate from the solar value of 5$\times$10$^{-17}$ s$^{-1}$ to the more extreme value of 10$^{-14}$ s$^{-1}$ have only very limited effects on the excitation diagram. The rather convincing fits to the H$_2$ data seem to indicate that these excitation effects by energetic photons could well be minimal in our case, but we repeat that their proper inclusion to our models is work in progress.

\end{appendix}

\end{document}